\documentclass{article}

\usepackage{PRIMEarxiv}

\usepackage[utf8]{inputenc} 
\usepackage[T1]{fontenc}    
\usepackage{hyperref}       
\usepackage{url}            
\usepackage{booktabs}       
\usepackage{amsfonts}       
\usepackage{nicefrac}       
\usepackage{microtype}      
\usepackage{lipsum}
\usepackage{fancyhdr}       
\usepackage{graphicx}       
\graphicspath{{figures/}}     

\usepackage{cite}
\usepackage{amsmath,amssymb,amsfonts}
\usepackage{algorithmic}
\usepackage{graphicx}
\usepackage{textcomp}

\usepackage{tikz, graphicx, mathtools, amsfonts, mathrsfs, amsmath, xfrac, amssymb, dsfont, enumitem,bbold, multirow}

\usepackage[ruled,vlined]{algorithm2e}
\SetKwInOut{Parameters}{Parameters}
\SetKwInOut{Initialization}{Initialization}

\DeclareMathOperator*{\argmax}{arg\,max}
\DeclareMathOperator*{\argmin}{arg\,min}

\newtheorem{remark}{Remark}
\newtheorem{lemma}{Lemma}
\newtheorem{proof}{Proof}

\newtheorem{assume}{Assumption}
  
\newcommand{\reals}{\mathbb{R}}

\newcommand{\diag}[1]{\mathrm{diag}\left\{#1\right\}}
\newcommand{\abs}[1]{\left\lvert#1\right\rvert}
\newcommand{\adjm}[1]{\mathrm{Adj}\left[#1\right]}
\newcommand{\degm}[1]{\mathrm{D}\left[#1\right]}
\newcommand{\norm}[1]{\left\lVert#1\right\rVert}

\newcommand{\ith}[1]{i^{\mathrm{th}}}

\title{Data-driven Cyberattack Synthesis against Network Control Systems
\thanks{This work has been submitted to IFAC for possible publication.\\
Omanshu Thapliyal is a PhD candidate with the School of Aeronautics \& Astronautics Engineering at Purdue University, West Lafayette, IN 47906. \\
Inseok Hwang is a professor at the School of Aeronautics \& Astronautics Engineering at Purdue University, West Lafayette, IN 47906.}}

\author{
  Omanshu Thapliyal\\
  Purdue University \\
  West Lafayette, IN USA\\
  \texttt{omanshu@purdue.edu} \\
   \And
  Inseok Hwang \\
  Purdue University \\
  West Lafayette, IN USA\\
  \texttt{ihwang@purdue.edu} \\
}

\begin{document}
\maketitle

\begin{abstract}
Network Control Systems (NCSs) pose unique vulnerabilities to cyberattacks due to a heavy reliance on communication channels.
These channels can be susceptible to eavesdropping, false data injection (FDI), and denial of service (DoS).
As a result, smarter cyberattacks can employ a combination of techniques to cause degradation of the considered NCS performance.
We consider a white-box cyberattack synthesis technique in which the attacker initially eavesdrops to gather system data, and constructs equivalent system model.
We utilize the equivalent model to synthesize hybrid cyberattacks -- a combination of FDI and DoS attacks against the NCS.
Reachable sets for the equivalent NCS model provide rapid, real-time directives towards selecting NCS agents to be attacked.
The devised method provides a significantly more realistic approach toward cyberattack synthesis against NCSs with unknown parameters.
We demonstrate the proposed method using a multi-aerial vehicle formation control scenario.
\end{abstract}


\section{Introduction}
Recent trends in combining networks with control systems allow control engineers to solve complex tasks, design sophisticated control schemes, and model cooperation of spatially separate entities, via data sharing across communication networks.
As a result, network control systems (NCSs) find varied applications in wireless sensor networks, industrial automation, distributed control systems, power grids, multi-robot cooperation, etc. \cite{sargolzaei2018security,gupta2009networked,yang2006networked}.
However, due to increased reliance on communication, often over insecure channels or in the presence of malicious entities, NCSs face increased cybersecurity threats compared to their centralized counterparts.
\cite{sargolzaei2018security} have noted numerous occurrences of such cyberattacks on NCSs: from the famous StuxNet, and the capturing of RQ-170 reconnaissance aircraft, to cybersecurity threats against autonomous driving systems.
Therefore, control theoretic methods can be utilized to inform NCS cybersecurity issues -- both, from the control designer and attacker perspectives.

\textit{Related Works:} In our previous work \cite{thapliyal2021learning}, we have classified cyberattacks under two categories: \textit{black-box} and \textit{white-box} cyberattacks.
Such a classification relies on the availability of NCS information to the attacker.
If the attacker has complete (or no) knowledge of the inner workings of the NCS, we refer to it as a white-box (or black-box) cyberattack, which are terms borrowed from the adversarial machine learning community \cite{kumar2020black,kalin2020black,zhang2020brute}.
While white-box techniques are more suited for assessing vulnerabilities of the control algorithms deployed on the NCSs, black-box techniques provide more realistic modeling methods for attack synthesis problems.

As such, white-box cyberattacks against NCS have been extensively covered in literature from the control designer's perspective, e.g., \cite{li2018false,yang2021risk,sargolzaei2019detection}.
\cite{li2018false} consider the attacker-defender interaction as a `leader-follower' game, where the eventual optimal responses from both players are compared with the Nash equilibrium solution.
\cite{yang2021risk} utilize a reachability based metric to quantify NCS security against stealthy injection attacks, where the attacker has complete system knowledge.
A Kalman filter augmented with a neural network is employed to detect false data injection (FDI) attacks an NCS by \cite{sargolzaei2018security}, while guaranteeing controller stability.
On the other hand, more realistic and sophisticated cyberattacks are often carried out as a combination of different attacks. 
Such white-box attacks often utilize statistical methods to develop system models \cite{kalin2020black,miao2021machine}, by exploiting well known communication vulnerabilities of NCSs (see \cite{sargolzaei2018security,liu2022reachability} for more detailed survey on NCS vulnerabilities, and FDI attack procedures, respectively).
As noted by \cite{thapliyal2021learning}, attackers can eavesdrop on black-box system data to develop auxiliary system models and perform injection attacks based on the model thus developed.
\cite{liu2022reachability} consider a different white-box FDI attack against a smart power grid (and the corresponding defense mechanism).
They assume the attacker is able to tap into communication channels to eavesdrop for power grid NCS measurements, and then carries out FDI attacks that are stealthy to chi-squared detection tests on the measurement residue.

However, most black-box attack methods require offline training \cite{miao2021machine}, computationally expensive machine learning \cite{thapliyal2021learning}, strong reachability assumptions \cite{liu2022reachability}, or complete knowledge of detection schemes \cite{yang2021risk}.
To this end, we would like to investigate the synthesis of white-box FDI attacks against NCSs, while including realistic attacker capabilities.
We assume the attacker is capable of observing the dynamical behavior of the NCS (in practice implemented by eavesdropping) to develop equivalent dynamical models.
We propose a dynamic mode decomposition (DMD) approach to obtain rapid real-time auxiliary approximations to the NCS dynamics.
The attack capabilities can be modeled as reachable sets of the auxiliary system, obtained in real-time through polytopic approximations \cite{hwang2005polytopic,varaiya2000reach}.
Additionally, the impact of the attack is demonstrated to be enhanced by isolating more vulnerable agents in the NCS.

\textit{Contributions:} In our work, we employ DMD combined with polytopic reachable set computation, to model the unknown NCS and attacker capabilities, respectively.
Therefore, the cyberattacks can be carried out with limited system data, and in real-time.
The agent isolation is posed as a repeated semi-definite program (SDP), thus allowing the attacker to carry out more directed attacks against the NCS, while only relying on the system data.
As a result, the proposed method provides a realistic attack synthesis scenario under practical attacker capabilities.

The remainder of this paper is organized as follows.
We present the problem formulation of cyberattack synthesis against the partially known network control system in Section \ref{sec2}.
We formulate a DMD-based scheme combined with reachable set estimation to devise the cyberattack synthesis methodology in Section \ref{sec3}.
In Section \ref{sec4}, we demonstrate the proposed method using a motivating scenario of injection attacks in actuator channels of a network of unmanned aerial vehicles (UAVs) engaging in formation flight and trajectory tracking.
Finally, we present our concluding remarks in Section \ref{sec5}.

\section{Problem Formulation}\label{sec2}
Consider a network control system (NCS) where the individual agents have a linear time-invariant (LTI) dynamics, under a coupled state feedback control law, as follows:
\begin{equation}\label{eq:lti-ncs}
\begin{split}
\dot{x}_i(t) &= A_i x_i(t) + B_i u_i(t),\\
u_i(t) &= K_{ii} x_i(t) + \sum_{j\in\mathcal N_i(t)} {K_{ij}\left(x_i(t)-x_j(t)\right)} 
\end{split}
\end{equation}
Here, the state of the $\ith{i}$ agent is given by $x_i\in\reals^n$, the coupled control input $u_i\in\reals^p$, the LTI system matrices $(A_i,B_i)$, and the state feedback gains $\{K_{ij}\}_{j=1}^N$, for a total of $N$ agents.
The agents in the NCS are connected according to an underlying graph $\mathcal G(t)=(\mathcal E(t), \mathcal V)$, a tuple of the node set $\mathcal V$ and the edge set $\mathcal E(t)\subseteq \mathcal V\times \mathcal V$.
Also, $\mathcal N_i(t)$ is the neighborhood of the $\ith{i}$  agent at time $t$, defined as the set $\{j:(i,j)\in\mathcal E(t),i\neq j\}$.
Let $\adjm{\mathcal G(t)}$ denote the graph adjacency matrix, and $\mathcal L_{\mathcal G(t)}$ the graph Laplacian, defined as:
\begin{equation}\label{eq:laplacian-def}
\begin{split}
\adjm{\mathcal G(t)} &\triangleq [a_{ij}] = 1 \text{ if } (i,j)\in\mathcal E(t), 0 \text{ otherwise} \\
\mathcal L_\mathcal G &\triangleq \degm{\mathcal G(t)} - \adjm{\mathcal G(t)}
\end{split}
\end{equation}
where $\degm{\mathcal G(t)}$ is the degree matrix with diagonal entries that denote the incoming degree (or $\abs{\mathcal N_i(t)}$) of the $\ith{i}$ agent.

From the attacker's perspective, the NCS dynamics can be rewritten as:
\begin{equation}\label{eq:stacked-lti-dynamics}
\dot{\pmb{x}}(t) \triangleq \mathbb A_{\mathcal G(t)} \pmb{x}(t)
\end{equation}
using the `stacked vector' definitions as:
\begin{equation}\label{eq:stacked-def}
\begin{split}
\pmb{x}(t) &\triangleq [x_1(t)^T,x_2(t)^T,\cdots,x_N(t)^T]^T\\
\mathbb A_{\mathcal G(t)} &\triangleq \diag{A_i+B_iK_{ii}}_{i\in\mathcal V} + B_i K_{ij}\otimes \mathcal L_{\mathcal{G}(t)}
\end{split}    
\end{equation}
The attacker intends to perform injection attacks to the system (\ref{eq:stacked-lti-dynamics}) with the following aims:
(a) cause safety violations/collisions between any two agents in the NCS, (b) while remaining undetected by the NCS monitoring protocols.
That is, false data injection (FDI) attacks to the system (\ref{eq:stacked-lti-dynamics}) can be written as $\dot{\pmb{x}}(t) = \mathbb A_{\mathcal G(t)} \pmb{x}(t) + \mathbb B^a \pmb{u}^a(t)$, for the injection attack vector $\pmb{u}^a(t)$, and a `vehicle selection matrix' $\mathbb B^a$.
The attack synthesis problem can then be written as:
\begin{equation}\label{eq:attacker-aim}
\begin{aligned}
\text{find } & \mathbb B^a, \pmb{u}^a(t)\\
\text{such that } & \dot{\pmb{x}}(t) = \mathbb A_{\mathcal G(t)} \pmb{x}(t) + \mathbb B^a \pmb{u}^a(t),  \\
& \exists\, t^*\geq t_1 \text{ where } \norm{x_i(t^*)-x_j(t^*)}_2\geq d^* ,\\
& \norm{\pmb{u}^a(t)}_2 \leq \rho \text{ for all $t\geq t_1$}
\end{aligned}  
\end{equation}
Here, $\rho$ denotes the FDI budget, and $d^*$ is some separation distance that the attacker wants to induce.
We make the following assumption for the FDI problem:
\begin{assume}\label{assume2}
The attacker is unaware of the system matrices $\left\{A_i(t),B_i(t),\{K_{ij}\}_{j\in\mathcal N_i(t)}\right\}_{i\in\mathcal V}$, but can collect discrete-time trajectory data $\{x_1(t),\cdots,x_N(t)\}_{t=t_0}^{t=t_1}$ over some time interval.
\end{assume}

In practice, the data collection in Assumption 1 is carried out via data association \& state estimation, or eavesdropping into communicated data that the agents use to formulate individual control laws.
Furthermore, severe cyberattacks are often preceded by simpler attacks to tap/eavesdrop on the system monitoring protocols or communication channels \cite{ding2018survey,thapliyal2021learning}.
As shown in Fig.~\ref{fig:schematic}, the attacker observes the state evolutions of the NCS agents.

\section{Methodology}\label{sec3}
To devise the injection attack, we first outline a method to infer the system matrix $\mathbb A_{\mathcal{G}(t)}$.
We will recap a Koopman operator based method (see works from  \cite{mauroy2020koopman,brunton2021modern} for more detail) that estimates underlying time-varying dynamical structure.

Note that the NCS in (\ref{eq:lti-ncs}) exhibits linear dynamics, but the stacked system in (\ref{eq:stacked-lti-dynamics}) is not perfectly linear due to the dependence on the time-varying graph $\mathcal G (t)$.
As a result, we employ a Koopman operator-based approach to identify the linear dynamics.
The Koopman operator can be thought of as a linear operator over measurable functional spaces that comes close (in operator norm) to the observed dynamics in the state space.
Consider the dynamics of some trajectory data collected from some dynamical map:
\begin{equation}\label{eq:dt-dynamics}
\pmb{x}_{k+1} = F(\pmb{x}_k)
\end{equation}
Even though the underlying dynamics can be nonlinear, the Koopman operator $\mathcal K$ acts on the space of all measurable functions $g:\mathcal X\to\mathcal X$ such that the evolution of these functions is linear.
That is, the trajectory of a function $g$ evolves linearly 
\begin{equation}
\begin{split}
g \circ F(\pmb{x}_{k}) &= \mathcal K \circ g(\pmb{x}_k)\Rightarrow g(\pmb{x}_{k+1}) = K\circ g(\pmb{x}_{k})
\end{split}
\end{equation}
under the action of $\mathcal K:\mathscr G\to\mathscr G$, where $\mathscr G$ is the space of \textit{observable functions}.

If the underlying state space is not finite, the Koopman operator, such that the observable trajectories are invariant in $\mathscr G$ and linear in $\mathcal K$, is infinite dimensional.
Almost all existing literature that relies on Koopman operator theory to perform system identification from state data, relies on finite dimensional approximations of $\mathcal K$ as a matrix $K$.
If we restrict ourselves to observable functions spanned by full-state measurements $\pmb{x}$, we try to find a matrix $K$ that approximates $\mathcal K$.
In reality, we carry out this approximation $K$ given \textit{data snapshots} as follows:

\begin{align}
\mathrm{X} &\triangleq \begin{bmatrix}
\lvert & \lvert & & \lvert \\
\pmb{x}_1 & \pmb{x}_2 & \cdots & \pmb{x}_w\\
\lvert & \lvert & & \lvert
\end{bmatrix},\,
\mathrm{X}^+ \triangleq \begin{bmatrix}
\lvert & \lvert & & \lvert \\
\pmb{x}_2 & \pmb{x}_3 & \cdots & \pmb{x}_{w+1}\\
\lvert & \lvert & & \lvert
\end{bmatrix}\\
K &= \argmin_{K} \norm{ \mathrm{X}^+ - K \mathrm{X}}_{F} = \mathrm{X}^+ \mathrm{X}^{\dagger}\label{eq:koop-dmd}
\end{align}
where $[\bullet]^\dagger$ denotes pseudoinverse and $\norm{\bullet}_{F}$ is the Frobenius norm.
Approximation in (\ref{eq:koop-dmd}) is referred to as \textit{dynamic mode decomposition} (DMD) \cite{mauroy2020koopman}, and can be thought of as linear regression given data snapshots.
The resulting approximation $K$ is a rank-$r$ (usually, $r < w$ for \textit{snapshot width} $w$) approximation of $\mathcal K$ confined to observables $g(\pmb{x_k})=\pmb{x}_k$.
Henceforth, we assume that we can obtain data snapshots of system (\ref{eq:lti-ncs}) by virtue of Assumption 1.
\begin{remark}
More complex observable families can be employed (e.g., Gaussian basis, polynomial basis, and neural networks to learn basis functions) depending on the degree of nonlinearity exhibited by the dynamical system, to result in extended dynamic mode decomposition methods. 
However, for our linear time-varying system identification, simple DMD suffices.
\end{remark}

Having solved for $K$, we have essentially performed regression for $\mathrm{X}^+\approx K \mathrm{X}$.
As a result, we can solve a modified version of the optimal control problem in (\ref{eq:attacker-aim}) as:
\begin{equation}\label{eq:attacker-modified}
\begin{aligned}
\text{find } & \mathbb B^a, \pmb{u}^a_k\\
\text{such that } & \dot{\pmb{x}} = K_k \pmb{x} + \mathbb B^a \pmb{u}^a,  \\
& \exists\, t^* \text{ where } \norm{x_{i}(t^*)-x_{j}(t^*)}_2\leq d^* ,\\
& \norm{\pmb{u}^a}_2 \leq \rho \text{ for all $k$}
\end{aligned}  
\end{equation}
To this end, we first define the set of all possible states that the agents in the NCS can take, when subjected to the norm-bounded injection attacks, as the \textit{reachable set} for agent $i$ as:
\begin{equation}
\mathcal R_i(\tau;\rho) \triangleq \left\{x_{i,k}:k\leq \tau, \norm{u^a}_{2}\leq \rho, \text{ and } \dot{\pmb{x}} = K_k \pmb{x}  \right\}
\end{equation}
Note that for our linearized system, reachable sets at some time $\tau$ are lent useful properties of convexity under system linearity. 
As a result, reachable sets of linear systems can be represented by propagating their boundaries or characterizations of their boundaries, e.g., interval representations in our previous work \cite{thapliyal2022approximate}, polytopic approximations by \cite{hwang2005polytopic,thapliyal2022approximating}, or ellipsoidal representations proposed by \cite{kurzhanski2000ellipsoidal}.

Next, we summarize the polytopic reachable set computation method for linear systems based on our work \cite{thapliyal2022approximating}, applied to the DMD approximation $\dot{\pmb{x}} = K_k \pmb{x}+\mathbb B \pmb{u}^a$, where $\pmb{u}^a(t)$ lies in a bounded set $\Omega \triangleq \{u\in\reals^{pN}:\norm{u}_{2}\leq \rho\}$.
We first bound $\Omega$ with an $s$-faced polytope, where the $\ith{i}$ face is given by:
\begin{equation}\label{eq:control-polytope}
\Omega\subseteq \bigcap_{i=1}^{s} {\{u\in\reals^{pN}\, \lvert\,\langle \nu_i, u\rangle \leq d_i\}}
\end{equation}
where $\nu_i$ is the normal vector, and $d_i$ the distance from the origin, for the $\ith{i}$ face.
Let $H(a,b)\triangleq\{x\in\reals^n:\langle a,x\rangle = b\}$ be the hyperplane with some normal vector $a$, at a distance $b$.
The evolution of the reachable set can then be expressed using the polytopic characterization of $\Omega$ from (\ref{eq:control-polytope}) as follows.
\begin{lemma}
Let hyperplanes $H_i(\lambda_{i,0},\gamma_{i,0})$ support the initial reachable set at time $t=0$ at points $x^*_{i,0}$ for each hyperplane $i$.
Then the reachable set at some time $\tau$, under a bounded input, $u^a$, can be represented as:
\begin{equation}
\begin{split}
\mathcal R(\tau;\rho) &\subseteq \bigcap^s_{i=1} H_i(\tau)\text{ where } H_i(\tau) \equiv (\lambda_i(\tau),\gamma_i(\tau))
\end{split}
\end{equation}
where the time-varying support points evolve according to $\dot{x^*_{i}}=K_kx^*_{i}+\mathbb B \pmb{u}^*$, and $\pmb{u}^*$ solves the optimal control problem $\argmax{\langle \lambda_i,  K_kx^*_{i}+\mathbb B \pmb{u}\rangle}$, and the costate evolution is given by $\dot{\lambda_i}=-K_k^T\lambda_i$.
The variable $\gamma_{i}(\tau)$ is the distance of the hyperplane $H_i$, i.e., $\langle \lambda_i(\tau), x^*_{i}(\tau)\rangle = \gamma_{i}(\tau)$
\end{lemma}

\begin{proof}
The proof follows from \cite{varaiya2000reach} and Theorem 1 in \cite{thapliyal2022approximating}. 
\end{proof}

Once the reachable set for the stacked system is obtained, $\mathcal R_i$ is obtained by projecting $\mathcal R$ on to the $\ith{i}$ state space coordinate.
Note that polytopic reachable sets are efficient to compute as the characterizations of reachable sets are a predefined number of hyperplanes.
Furthermore, hyperplanes are characterized by points of contact $x^*_i$ and normal vectors $\lambda_i$, which evolve under the linear dynamics.
This evolution takes place under the optimal control law $\pmb{u}^*_i$, which is also easy to compute for the polytopic set $\Omega$.
This is because the maximum is guaranteed to occur on one of the vertices of the polytope in (\ref{eq:control-polytope}) that bounds $\Omega$ \cite{varaiya2000reach}.

\begin{figure}[!t]
    \centering
    \includegraphics[width=0.75\columnwidth]{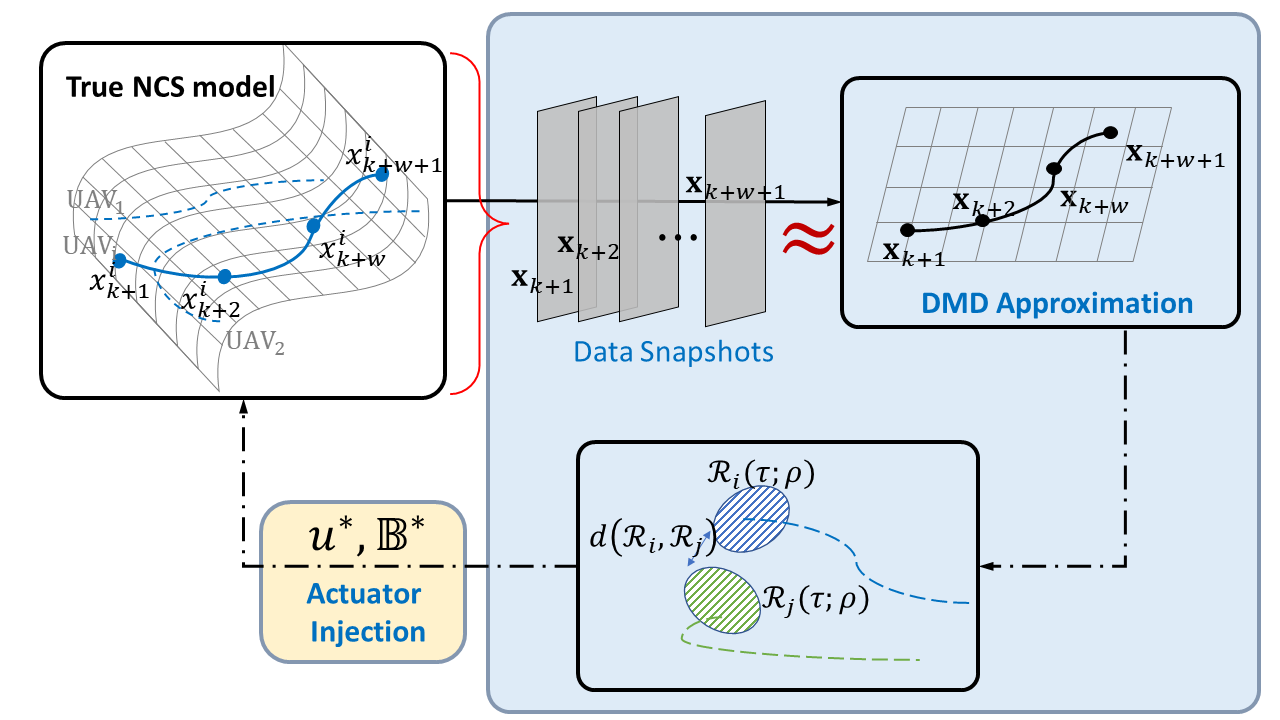}
    \caption{Schematic of the proposed cyberattack synthesis method}
    \label{fig:schematic}
\end{figure}

Once $\mathcal R_i(\tau;\rho)$ are computed for some time $\tau$, (\ref{eq:attacker-modified}) is solved as follows.
The attacker chooses $\mathbb B$ to indicate the agents being attacked (e.g., if the $\ith{i}$ agent is being attacked, $\mathbb B$ has an identity in the $\ith{i}$ place, and zero matrices for the remaining agents).
The FDI attack vector $\pmb{u}^*$ that solves $d(\mathcal R_i(\tau+\Delta t),\mathcal R_j(\tau+\Delta t))\geq d^*$ for some $\Delta t$, also solves (\ref{eq:attacker-modified}) for $t^*\in[\tau,\tau+\Delta t]$.
This is because the FDI attack of $\pmb{u}^*$ pushes the reachable sets for agents $i$ and $j$ at least $d^*$ units apart in some time $\Delta t$, which necessarily means that the constraints in (\ref{eq:attacker-aim}) are automatically satisfied.
As a result, the reachable sets thus found encode the attacker's aim in the original problem.
A detailed application of the proposed method is presented in the following section, and a summary of the method is summarized in Fig.~\ref{fig:schematic}.

\section{False Data Injection Attack against Formation Control of UAVs}\label{sec4}
In this section, we consider a network of 5-UAVs performing formation control while trying to follow a desired trajectory.
Such problems often arise in the NCS literature, NCS cybersecurity, distributed control, and UAV path planning problems \cite{kwon2018sensing}.

\subsection{NCS Model}
\begin{figure}[!ht]
    \centering
    \resizebox{0.4\columnwidth}{!}
    {
    \input{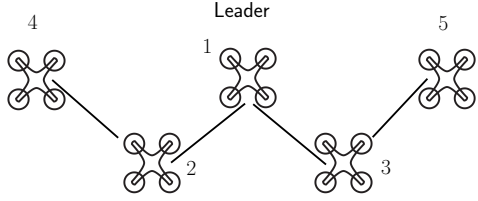}
    }
    \caption{Desired UAV NCS formation}
    \label{fig:formation}
\end{figure}

\begin{figure}[!t]
    \centering
    \includegraphics[width=0.7\columnwidth]{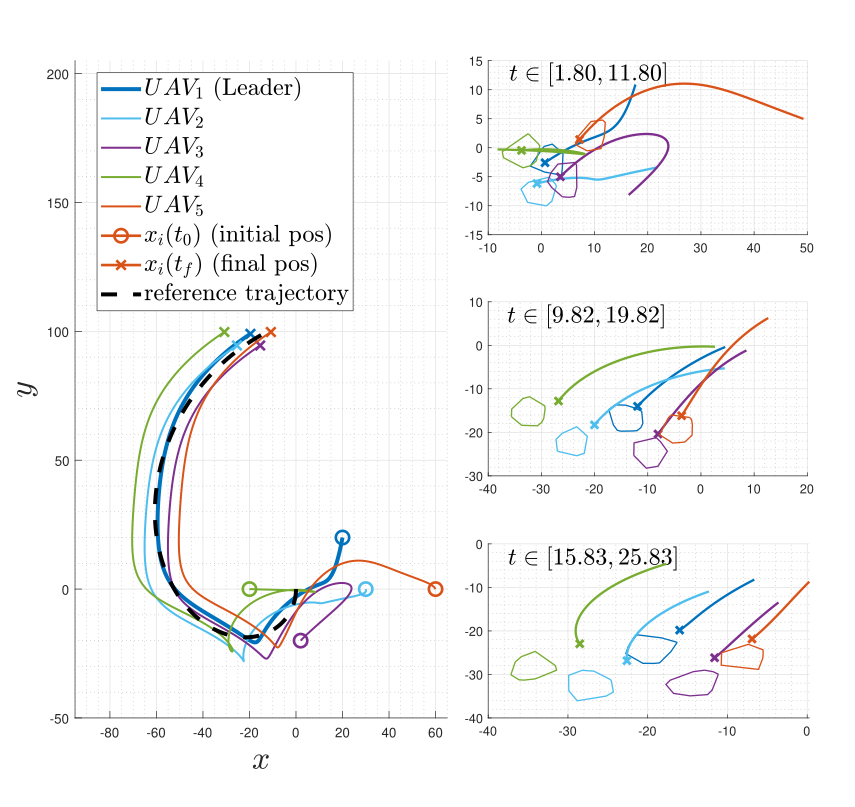}
    \caption{Formation control for 5-UAV network under gains $K_{ij}$ \textit{(left)}; Approximate reachable sets for DMD-based auxiliary model \textit{(right)}}
    \label{fig:nominal-system}
\end{figure}

Although the proposed method is developed for continuous time NCSs, the numerical implementation has to be carried out in discrete time.
The $\ith{i}$ UAV in the NCS has its dynamics governed by the discrete-time version of (\ref{eq:lti-ncs}) as $x_{i,k+1}=A_ix_{i,k}+B_iu_{i,k}$, with the system matrices:
\begin{equation*}
A_i = \begin{bmatrix}    
1 & \Delta t & 0 & 0\\
0 & 1 & 0 & 0\\
0 & 0 & 1 & \Delta t\\
0 & 0 & 0 & 1
\end{bmatrix},\;
B_i = \begin{bmatrix}    
\sfrac{\Delta t^2}{2} & 0\\
\Delta t & 0\\
0 & \sfrac{\Delta t^2}{2}\\
0 & \Delta t
\end{bmatrix}
\end{equation*}
The dynamics above are the general double integrator in the 2D plane, and the $\ith{i}$ UAV's state is defined as $x_i=[x, \dot{x},y,\dot{y}]^T$.
The formation control problem for the UAVs is solved by the following distributed control input:
\begin{align}
u_{1,k} &= K_{1}(x_{1,k}-x^\star_k)+\sum_{j\in\mathcal N_1}K_{1j}(x_{1,k}-x_{j,k}-x^\star_{1j}) \label{eq:formation-control-input-leader} \\ 
u_{i,k} &= \sum_{j\in\mathcal N_i}K_{ij}(x_{i,k}-x_{j,k}-x^\star_{ij}),\;i=\{2\cdots,5\} \label{eq:formation-control-input}
\end{align}
Equation (\ref{eq:formation-control-input-leader}) denotes the control input of the leader UAV, according to the reference trajectory $x^\star_k$ which the entire formation must follow.
Equation (\ref{eq:formation-control-input}) denotes the control sequence for all of the remaining follower UAVs who try to conform to a prescribed desired formation $x_{ij}^\star$.
Note that for the UAV NCS described by (\ref{eq:formation-control-input-leader}) and (\ref{eq:formation-control-input}), the NCS error dynamics follows (\ref{eq:stacked-def}), as noted by \cite{kwon2018sensing}.
The desired formation trajectory is given by $x^\star_k=[-k\sin(\sfrac{3k}{100}), 1, -k\cos(\sfrac{3k}{100}),1]^T$, for a simulation time of $100$s, a sampling time of $\Delta t= 0.2$s, and a desired formation as shown in Fig.~\ref{fig:formation}.
The stabilizing gain to guarantee formation control was found by \cite{kwon2018sensing} to be:
\begin{equation*}
K_{ij}=\begin{bmatrix}
-0.2263 & -0.4712 & 0 & 0\\
0 & 0 & -0.2263 & -0.4712
\end{bmatrix}
\end{equation*}
The resulting trajectory of the 5-UAV NCS is shown in Fig.~\ref{fig:nominal-system}~(left).
The initial positions of the UAVs, denoted by circles, were chosen randomly. 
The final positions denoted by crosses confirm a successful formation and trajectory tracking by the leader UAV.

\subsection{Attack Synthesis}
In order to find the auxiliary model $\pmb{x}_{k+1}\approx K \pmb{x}_{k}$ using DMD, it is assumed that the attacker performs state estimation to obtain the data matrices $\mathrm{X}$ and $\mathrm{X}^+$ in (\ref{eq:koop-dmd}), with a snapshot width $w=50$, at a sampling rate of $\Delta t=0.2$s.
The FDI attack budget is given by $\rho = 0.05$.
We approximate this admissible attack set as an 8-sided polygon with vertices chosen randomly such that the $\norm{\pmb{u^a}}\leq \rho$ is circumscribed by the polygon.
The attacker then computes the reachable sets according to Lemma 1, and propagates them over time to synthesize FDI attacks.
The propagated reachable sets at selected time instances are shown in Fig.~\ref{fig:nominal-system}~(right).
The tails in the comet plots denote the previous trajectory of the given UAV over the intervals shown.

Since UAVs are trying to conform to a specified formation, attacked agents are selected based on the reachable sets at a given time $\tau$ as:
\begin{align}
(i^*,j^*) &=\argmax_{i,j}{d\left(\mathcal R_i(\tau;\rho), \mathcal R_j(\tau;\rho)\right)} \label{eq:agent-select}\\
\pmb{u}^a &= \argmax_{\norm{u}\leq 0.05} {d\left(\mathcal R_i(\tau+\Delta t;\rho), \mathcal R_j(\tau+\Delta t;\rho)\right)} \label{eq:fdi-attack}
\end{align}
where the UAVs to be attacked $(i^*,j^*)$ are chosen whose reachable sets are farthest apart (\ref{eq:agent-select}), and the FDI attack is chosen to drive them further apart at the next time step, while being subject to the attack budget.
These distances can be computed efficiently as the sets are polytopic.

\begin{figure}[!t]
    \centering
    \includegraphics[width=0.7\columnwidth]{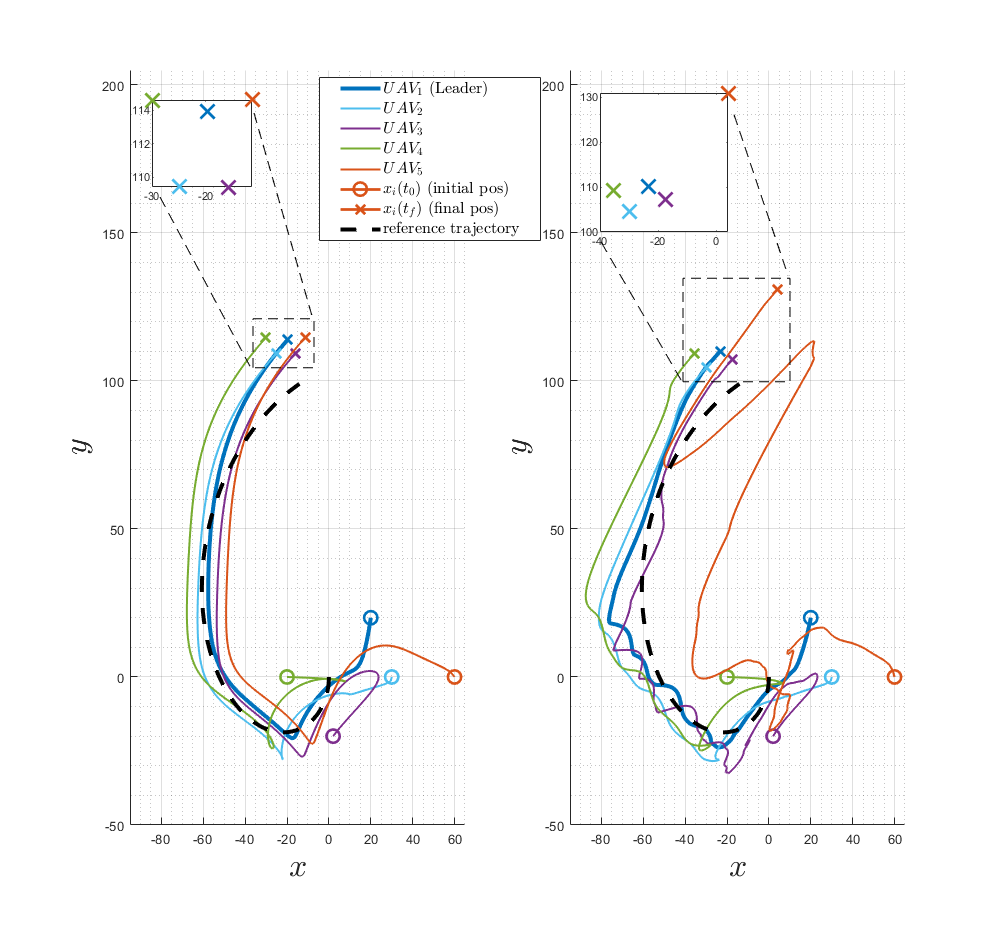}
    \caption{Impact of false data injection attacks on UAV trajectories: FDI attacks \textit{(left)}; FDI combined with DoS on agent 5, link 5--3 \textit{(right)}}
    \label{fig:attacked-system}
\end{figure}

The targeted agents change over time, depending on the relative locations of their reachable sets, as in Fig.~\ref{fig:schematic}.
Finally, $\mathbb B^a$ is the corresponding matrix with a zero matrix at all positions, and an identity matrix at positions $(i^*,j^*)$ to denote FDI attacks taking place at the selected agents.
The effect of such an attack over time can be observed in the trajectories shown in Fig.~\ref{fig:attacked-system}~(left).
Note that the stacked system evolves under $\mathbb A_{\mathcal G}$, and the formation error is guaranteed to go to zero under the proposed gain, as the dynamics of the stacked error is stable \cite{kwon2018sensing}.
As a result, the FDI attack can be seen to cause a disruption in only the trajectory tracking error, and the desired formation is still attained (see Fig.~\ref{fig:attacked-system}~(left)).

To cause a disruption in the formation, the stability of the stacked system can be compromised if the underlying connected graph can be disconnected.
This is carried out by preceding the FDI attack with a sustained denial of service (DoS) attack on a `vulnerable agent', whose communication link when disrupted causes the underlying graph to be disconnected.
However, the auxiliary DMD system evolves under a time-varying matrix $K$, which does not necessarily preserve the underlying graph structure in (\ref{eq:stacked-def}).
This is because the matrix $K$ was obtained by minimizing the trajectory error in (\ref{eq:koop-dmd}), and does not necessarily find the matrix $\mathcal L_{\mathcal G}$ (as DMD can be though of as a special case of L$_2$ regression \cite{mauroy2020koopman}).
Therefore, to find an estimate of the underlying graph Laplacian, the attacker solves the following equation:
\begin{equation}\label{eq:laplacian-min}
\begin{split}
\hat{\mathcal L} &= \argmin_{L} \norm{K - \left(S + T\otimes L \right)}_F\\
\end{split}    
\end{equation}
for some matrices $S$ and $T$, such that $L$ is a candidate Laplacian, i.e., $L$ satisfies $L\succeq 0$ and $L\mathbb 1 = 0$.
The minimization of the Kronecker product above can be rewritten as a minimization over $\gamma$ for some bound $\left(K - \left(S + T\otimes L \right)\right)^T\left(K - \left(S + T\otimes L \right)\right)\preceq \gamma^2 I$.
This can be rewritten as a `quasi' semi-definite program (SDP) using Schur complement as:
\begin{equation}\label{eq:sdp}
\begin{split}
& \min{\gamma}\\
& \text{subject to } \begin{bmatrix}
\gamma I & K - \left(S + T\otimes L \right)\\
\left[K - \left(S + T\otimes L \right)\right]^T & \gamma I
\end{bmatrix} \succ 0\\
& L\succeq 0, L\mathbb 1 = 0
\end{split}    
\end{equation}
Note that the equation above is a proper SDP if the matrix $T$ is fixed.
Similar SDP formulations to minimize matrix norms over Kronecker product are proposed by \cite{dressler2022kronecker}.
We can solve (\ref{eq:sdp}) as an alternating SDP as follows.
Starting with an arbitrary but fixed value of $S$ and $T$, we solve the SDP in (\ref{eq:sdp}) to find $L$.
Next, fix $T$ and the value $L$ thus obtained to solve for $S$, followed by solving for $T$ after fixing $L$ and $S$, respectively.
The alternating SDP procedure is carried out until $\gamma$ stops improving above a prefixed threshold, resulting in $\hat{\mathcal L}$.
This alternating SDP procedure is derived by \cite{dressler2022kronecker} in more detail.
\begin{figure}[!t]
    \centering
    \includegraphics[width=0.7\columnwidth]{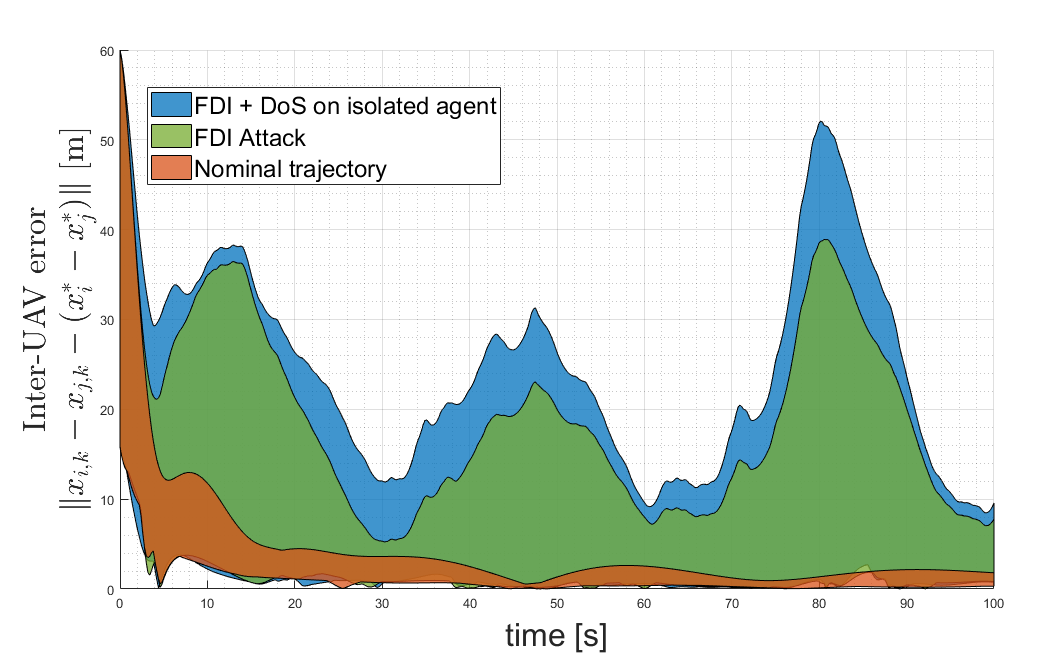}
    \caption{Inter-UAV errors plotted against time}
    \label{fig:attacked-system-cumulative-error}
\end{figure}
Using this alternating SDP method, the attacker was able to recover the graph Laplacian matrix $\hat{\mathcal L}$ in 16 iterations of the alternating SDP where $\gamma$ stopped improving beyond $10^{-6}$.
Next, solving for the second largest eigenvalue of the Laplacian $\lambda_2(\hat{\mathcal L})$ immediately provides the vulnerable nodes in the underlying strongly connected graph (i.e., there exist connecting paths between any two arbitrary nodes of the graph).
This is because the second largest eigenvalue of the graph Laplacian is an immediate metric of graph connectivity \cite{west2001introduction}, and all distributed control of the form (\ref{eq:lti-ncs}) relies on the underlying graph being strongly connected \cite{clarke2022attack}.
The eigenvector corresponding to $\lambda_2(\mathcal L_{\mathcal G})$, called Fiedler eigenvector, provides immediate relative importance of each node of the graph $\mathcal G$ towards graph connectivity \cite{west2001introduction}.
As a result, the attacker chooses to DoS the $\ith{5}$ UAV (link 5--3), thereby causing the underlying graph to be disconnected, and the stacked system is no longer stable.
The FDI attacks are carried out as outlined earlier.

The resulting trajectory can be seen in Fig.~\ref{fig:attacked-system}~(right).
Compared to the reachable set-based FDI attacks, the attack mechanism of DoS attack on UAV 5 followed by FDI attacks causes disruption in formation as well as a failure in trajectory tracking.
The same is observed in the plot of inter-UAV position errors, $\Vert x_{i,k} - x_{j,k} - (x^*_i - x_j^*)\Vert$, shown for the nominal NCS, FDI attacks on the NCS, and the final FDI combined with alternating SDP-based DoS attacks to isolate UAV 5 (see Fig.~\ref{fig:attacked-system-cumulative-error}).
As a result, the attacker can cause the formation errors to increase and accumulate over time.
Therefore, by utilizing the reachable sets, the attacker can cause disruption in formation control and trajectory tracking in the UAV NCS discussed.

\section{Conclusion}\label{sec5}
In this paper, we developed a novel cyberattack synthesis mechanism targeting network control systems (NCSs) with unknown dynamics.
We proposed a dynamic mode decomposition-based method to first estimate reachable sets of the NCS agents, followed by false data injection (FDI) attacks by driving the reachable sets as far apart as possible.
We demonstrated the proposed method using an illustrative scenario of unmanned aerial vehicle (UAV) formation flight and trajectory tracking.
The proposed method was observed to cause failure in both, formation and trajectory tracking, upon preceding the FDI attacks by denial of service (DoS) attacks to isolate certain agents.

Future work will be to utilize exact reachable sets from the defender's perspective to provide guarantees on the controller against FDI attacks of a fixed budget.

\section*{Acknowledgment}
{The authors would like to acknowledge that this work was supported by NASA University Leadership Initiative (ULI) under Grant 80NSSC20M0161.}
\bibliographystyle{plain}  

\end{document}